\documentclass[12pt]{article}
\usepackage{amssymb,amsmath,epsfig}

\begin{document}

\title{\bf  Dynamics of Charged Bulk Viscous Collapsing Cylindrical Source With Heat Flux}
\author{S. M. Shah \thanks{syedmunawarshah71@hotmail.com} and  G. Abbas \thanks{ghulamabbas@iub.edu.pk}
\\Department of Mathematics, The Islamia University \\of Bahawalpur, Bahawalpur-63100, Pakistan.}
\date{}
\maketitle
\begin{abstract}
 In this paper, we have explored the effects of dissipation on the dynamics of charged bulk viscous collapsing cylindrical source which allows the out follow of heat flux in the form of radiations. Misner-Sharp formulism has been implemented to drive the dynamical equation in term of proper time and radial derivatives. We have investigated the effects of charge and bulk viscosity on the dynamics of collapsing cylinder. To determine the effects of radial heat flux, we have formulated the heat transport equations in the context of M$\ddot{u}$ller-Israel-Stewart theory by assuming that thermodynamics viscous/heat coupling coefficients can be neglected within some approximations. In our discussion, we have introduced the viscosity by the standard (non-casual) thermodynamics approach. The dynamical equations have been coupled with the heat transport equation equation, the consequences of resulting coupled heat equation have been analyzed in detail.
\end{abstract}
\section{Introduction}
The stars are composition of some nuclear matter which is continuously gravitating and is attracted towards its center due to gravitational
interaction of its particles. This phenomena, in theory of general reactivity is known as gravitational collapse. The description about this
phenomena is the main objective of the relativistic theories of gravity (including general relativity) \cite{1,2}. Oppenheimer and Snyder \cite{3} theoretically illustrated the process of collapse in 1939, they observed the contraction of highly idealized spherically symmetric dust cloud. They used exterior and interior spacetimes as Schwarzschild metric and Friedman like solution respectively. An enormous contribution in the research of gravitational collapse has been added by Vaidya \cite{4}, who provide the exterior gravitational field of a stellar body, giving out radiations. Misner and Sharp \cite{5,6} studied perfect fluid spherically symmetric collapse and also some authors \cite{8}-\cite{14} considered it in different situation.

Rossland \cite{15} proved that the atoms are converted into ions with great strength and the law of central force should be observed by the forces between the free particles, its order of magnitude should be greater than that of remaining forces acting between neutral atoms. The effect of electrical forces is fairly large if the star is built of heavy elements with 1.5 times solar mass and mean molecular weight 2.8 unit. Eddington \cite{16} explored that in the internal electrical field of star, the electric potential $\phi$ directly relates the gravitational potential $\psi$, the mass $m$ and charge $e$ of a proton and a scalar parameter $\alpha$ affected by density $n_{i}$ of the ions, atomic weight $A_{i}$ of the ions and the effective charge $eZ_{i}$. Mitra \cite{17a} introduced the fact that the formation and evolution of stars would happened due to gravitational collapse, which is high energy dissipating process and can be characterized into two respective cases, free Streaming approximation and diffusion approximation. In the free streaming approximation case, Tewari added some models \cite{17}-\cite{19} by the solution of Einstein field equation with different approach. A number of distinguish researchers such as: Bonner et al.\cite{20}, Bowers and Liang \cite{21}, de Oliveria \cite{22}, Mahraj and Govender \cite{23}, Ivanov \cite{24} and Phinheiro and Chan \cite{25} discussed many realistic models in diffusion approximation with anisotropy, inhomogeneity, viscosity, electromagnetic field and also observed different dissipative processes analytically.

Since then, a huge amount of literature \cite{8}-\cite{14} on gravitational collapse consider the spherical symmetry of the star which is simplest geometry. In order to determine a realistic model of gravitational collapse, it would be interesting to study the dynamics of collapsing star with non-spherical background. It would be implied by the existence of  gravitational waves that cylindrical and plane symmetries are more important for non-spherical back ground. The cylindrical sources may serve a test bed for numerical relativity, quantum gravity, and for probing cosmic censorship and hoop conjecture, among other important issues, and represent a natural tool to seek the physics that lies behind the two independent parameters in Levi-Civita metric \cite{L1}. Herrera et al.\cite{26} have discussed gravitational collapse and junction/interface criteria for gravitating source which has cylindrical geometry. Sharif and Ahmad \cite{27} have predicted that gravitational radiations can be emitted during gravitational collapse of two perfect fluids. Di Prisco et al.\cite{28} studied the shear-free conditions and cylindrical gravitational waves by taking the Einstein-Rosen spacetime in the exterior of general non-static cylindrical spacetime. Nakao and Morisawa  \cite{29} have explored the gravitational radiations from the collapse of a hollow cylinder.

 Since Einstein and Rosen \cite{31} initially predicted cylindrical gravitational waves theocratically, the observational evidences of
gravitational waves through advance detectors such as LIGO \cite{34} and GEO \cite{36} have motivated the researchers to study the cylindrically symmetric gravitating source. The formation of naked singularity during the generic gravitational collapse would be expected during cylindrical gravitational collapse. Several numerical approximations \cite{37} depicts  the emission of
gravitational cylindrical waves from cylindrical gravitating source. These results have
been verified analytically by Nakao and Morisawa \cite{40}. \textit{During the recent years, many attempts \cite{41}-\cite{43} have been made to study the dynamics of collapsing cylindrical sources but all these involves the cylindrical spacetimes  which are very similar to spherical spacetimes. In the current study, we have taken the nontrivial cylindrical spacetime.}

In the present study, we have considered the two types of dissipation processes, heat dissipation associated to the radial heat flux and bulk viscosity. These both dissipative terms have been included in the stress energy tensor of the gravitating source. In order to see the effects of these terms on the dynamics of the collapse, we used the heat transport equations in the context of M$\ddot{u}$ller-Israel-Stewart theory \cite{I1}-\cite{I3}. Such equations provide the physically reasonable heat transportation process as compared to Landau-Eckart approach \cite{I9,I10}(by neglecting the thermodynamics viscous/heat coupling coefficients). The bulk viscosity has been described according to standard (non-casual) irreversible thermodynamics approach in the stress energy tensor of the gravitating source. The inclusion of the bulk viscosity in the fluid implies that we are assuming the relativistic Stokes equations, which corresponds to the irreversible thermodynamics. This equation does not satisfy the casuality, because implicitly it is assumed that corresponding bulk viscosity relaxation times vanishes, this assumption is valid within some approximations.

The plan of the paper is as follows: In section \textbf{2} we present cylindrical source and field equations. The dynamical equations with Misner-Sharp approach have been presented in section \textbf{3}. The derivation of heat transport equation and its coupling is given section \textbf{4}. The last section is devoted to summery of results of this paper.

\section{Gravitating Source and Field Equations }
In this section, we shall briefly introduce matter source, geometry of star for both interior and exterior regions and the field equations for the
charged radiating bulk viscous source. The cylindrically symmetric  spacetime \cite{azm} is
\begin{eqnarray}\label{a1}
ds^{2}_{-}=-X^2(r,t)dt^{2}+Y^2(r,t)dr^{2}+R^{2}(r,t)d\theta^{2}+dz^{2},
\end{eqnarray}
where $-\infty\leq t\leq\infty$, $0\leq r$, $-\infty\leq z\leq\infty$, $0\leq \theta\leq 2\pi$.

Inside the cylindrical star, we take charged, anisotropic, bulk viscous fluid with radial heat flux, which has the following form of energy momentum tensor
\begin{eqnarray}\nonumber
T_{\alpha\beta}&=&(\mu+P_{r})V_{\alpha}V_{\beta}-(P_{r}-P_{z})S_{\alpha}S_{\beta}+(P_{r}-P_{\theta})\chi_{\alpha}\chi_{\beta}-(g_{\alpha\beta}+
V_{\alpha}V_{\beta})\xi\Theta\\\label{3}
&+&q_{\alpha}V_{\beta}+V_{\alpha}q_{\beta}+P_{r}g_{\alpha\beta}
+\frac{1}{4\pi}\left(F^{\gamma}_{\alpha}F_{\beta\gamma}-
\frac{1}{4}F^{\gamma\delta}F_{\gamma\delta}g_{\alpha\beta}\right),
\end{eqnarray}
where $\mu$ is energy density, $P_{r}$ is pressure perpendicular to $z$ direction,  $P_{\theta}$ is pressure in $\theta$ direction, $P_{z}$ is pressure in $z$ direction, $V_{\alpha}$ is four velocity, $\xi$ is coefficient of bulk viscosity, $\Theta$ is expansion scalar and $q_{\alpha}$ is radial heat flux. Also,
$F_{\alpha\beta}=-\phi_{\alpha,\beta}+\phi_{\beta,\alpha}$ is the
Maxwell field tensor with four-potential is $\phi_\alpha$. Moreover,
$S_{\alpha}$ and $\chi_{\alpha}$ are the unit four-vectors which
satisfy the following relations:
\begin{equation*}
\chi^{\alpha}\chi_{\alpha}=S^{\alpha}S_{\alpha}=1,\quad
V^{\alpha}V_{\alpha}=-1,\quad
V^{\alpha}S_{\alpha}=S^{\alpha}\chi_{\alpha}=V^{\alpha}\chi_{\alpha}=0.
\end{equation*}
The four vector velocity $V_{\alpha}$ and four vectors $\chi_{\alpha}$ and $S_{\alpha}$ can be defined as follows
\begin{equation*}\label{3}
\chi_{\alpha}=R{\delta}^{2}_{\alpha},\quad
v_{\alpha}=-X\delta^{0}_{\alpha},\quad
S_{\alpha}={\delta}^{3}_{\alpha}.
\end{equation*}

The Maxwell field equations are
\begin{equation}\label{a2}
F^{\alpha\beta}_{~~;\beta}={4\pi}J^{\alpha},\quad
F_{[\alpha\beta;\gamma]}=0,
\end{equation}
where $J_\alpha$ is the four-current. It is assumed that there exists only non-vanishing electric scalar potential and magnetic vector potential will be zero then four potential takes the following form
\begin{equation*}
\phi_{\alpha}={\phi}{\delta^{0}_{\alpha}},\quad
J^{\alpha}={\zeta}V^{\alpha},
\end{equation*}
where $\zeta(r,t)$ is charge density and $\phi(r,t)$ is electric scalar potential.
\newline

The expansion scalar is
\begin{eqnarray}\label{a3}
 \Theta&=&\frac{1}{X}\left(\frac{2\dot{Y}}{Y}+\frac{\dot{R}}{R}\right),
 \end{eqnarray}
where dot and prime denote differentiation with respect to $t$ and
$r$ respectively.

The set of Einstein-Maxwell field equations is
\begin{eqnarray}\label{d}
\kappa\left(\mu-\frac{\pi}{2}E^{2}\right)X^{2}&=&\frac{{\dot{Y}}\dot{R}}{YR}+\left(\frac{X}{Y}\right)^{2}\left(\frac{X'R'}{XR}-\frac{R''}{R}\right),
\\\label{q}
\kappa qXY^{2}&=&\frac{{\dot{R'}}}{R}-\frac{\dot{Y}R'}{YR}-\frac{\dot{R}X'}{RX},
\\\label{pr}
\kappa\left(P_{r}-\xi\Theta+\frac{\pi}{2}E^{2}\right)Y^{2}&=&\frac{X'R'}{XR}+\left(\frac{Y}{X}\right)^{2}\left(-\frac{\ddot{R}}{R}+\frac{\dot{X}\dot{R}}{XR}\right),
\\\label{pt}
\kappa\left(P_{\theta}-\xi\Theta-\frac{\pi}{2}E^{2}\right)&=&\left(\frac{1}{XY}\right)\left(\frac{\dot{X}\dot{Y}}{X^{2}}-\frac{X'Y'}{Y^{2}}-\frac{\ddot{Y}}{X}+\frac{X''}{Y}\right),\\\nonumber
\kappa\left(P_{z}-\xi\Theta-\frac{\pi}{2}E^{2}\right)&=&-\frac{\ddot{Y}}{X^{2}Y}+\frac{X''}{XY^{2}}-\frac{\ddot{R}}{X^{2}R}
-\frac{X'Y'}{XY^{3}}
+\frac{\dot{X}}{X^{3}}\left(\frac{\dot{R}}{R}+\frac{\dot{X}}{X}\right)\\ &-&\frac{R'}{Y^{2}R}\left(\frac{Y'}{Y}+\frac{X'}{X}\right)
-\frac{\dot{Y}\dot{R}}{X^{2}YR}+\frac{R''}{Y^{2}R},
\end{eqnarray}
 where $E(r,t)=\frac{\hat{Q}(r)}{2\pi R}$ with total charge $\hat{Q}(r)=4\pi\int^r_{0}{\zeta}{YR}dr$.

Analogous to Misner-Sharp mass in spherically symmetry, Thorne \cite{31} introduced the mass function for cylindrical spacetime in term of gravitational
C-energy per unit length of the cylinder. The specific energy $m(r,t)$ in the presence of electromagnetic field is \cite{zaeem}
\begin{equation}\label{m}
m\left(r,t\right)=\frac{l}{8} \left[1+\left(\frac{\dot{R}}{X}\right)^{2}-\left(\frac{R'}{Y}\right)^{2}\right]+\frac{\hat{Q}^{2}l^2}{2R}.
\end{equation}
Here $l$ is the constant specific length of the cylinder.

Let ${\Sigma}$ is a boundary surface, which separates the interior region (defined in Eq.(\ref{a1})) from the exterior region, the exterior region is described by cylindrically symmetric manifold in the retarted time coordinate as \cite{azm}
\begin{equation}\label{ex}
ds^{2}_{+}=-\left(-\frac{2M(\nu)}{\tilde{R}}+\frac{\tilde{q}^{2}(\nu)}{\tilde{R}^{2}}\right)d\nu^{2}-2d\nu d\tilde{R} + \tilde{R}^{2}(d\theta^{2}+\gamma^{2}dZ^2),
\end{equation}
where $M(\nu)$ and ${\tilde{q}(\nu)}$ are mass and charge, respectively and $\gamma^{2}=-\frac{\Lambda}{3}$, $\Lambda$ is cosmological constant.
Using the continuity of line elements and extrinsic curvature of line elements given by Eqs.({\ref{a1}) and ({\ref{ex}) and field equations, we get  \cite{zaeem}
\begin{equation}
P_r-\xi\Theta=^{\Sigma}(qY), \quad\quad m-M=^{\Sigma}\frac{l}{8} , \quad\quad {\hat{Q}^{2}l^2} =^{\Sigma}{\tilde{q}^{2}}, \quad\quad l =^{\Sigma}4\tilde{R}.
 \end{equation}
These are the necessary conditions for the smooth matching of internal and external geometries of cylindrical stars over the hypersurface ${\Sigma}$. For the assumed cylindrical source the difference of $M$ and $m$ (specific energy) is non-zero in general and constraint $l =^{\Sigma}4\tilde{R}$, must be satisfied over ${\Sigma}$.

\section{Dynamical Equations}

According to Misner and Sharp \cite{5,6}, we introduce the proper time derivative $D_{t}$ as follows
\begin{equation}\label{92}
D_{t}=\frac{1}{X}\frac{\partial}{\partial t}.
\end{equation}
The velocity $U$ of the fluid collapse may be stated in terms of Eq.(\ref{92}) as
\begin{equation}\label{m1}
U=D_{t}R<0\quad\quad\quad ({in~ case~of ~collapse}).
\end{equation}
Hence Eq.(\ref{m}), yields
\begin{equation}\label{m1}
\frac{\acute{R}}{Y}=\left(1+U^{2}-\frac{8m}{l}+\frac{4\hat{Q}^{2}l}{R}\right)^{\frac{1}{2}}=\hat{E},
\end{equation}
where $\hat{E}$ is the energy of an element of the fluid that undergoes collapse.
The proper time derivative of mass function described in Eq.(\ref{m}), takes the following form
\begin{equation}\label{m0}
D_{t}m=l\left(\frac{\dot{R}\ddot{R}}{4X^{3}}-\frac{\dot{R^{2}}\dot{X}}{YX^{4}}-\frac{R'\dot{R'}}{4Y^{2}X}+\frac{R'^{2}\dot{Y}}{4XY^{3}}\right)
-\frac{\dot{R}\hat{Q}^{2}l^2}{2XR^{2}}.
\end{equation}
Using Eqs.(\ref{d}) and (\ref{pr}) and $E=\frac{\hat{Q}}{2\pi R}$, we obtain
\begin{equation}\label{m00}
D_{t}m=-2\pi l\left(\hat{E}qB+U(P_{r}-\xi\Theta)\right)R.
\end{equation}
This equations provide the rate of change of total energy available inside the cylinder of radius $R$.
Here, we briefly explain the effect of each term on the change of total internal energy, on the right hand side of above equation
the term $\hat{E}qB$ being the multiple of negative sign implies the amount of heat energy leaving the surface of cylindrical star. In other words
out flow of heat from the collapsing system reduces the total energy of the system.
In the second term $U(P_{r}-\xi\Theta)<0$, (as $U<0,$ and $\Theta<0$ due to collapse and $\xi>0$), hence this having pre-factor $-2\pi$,
increases the energy inside the collapsing source.
The proper radial derivative $D_{R}$ is used to observe the dynamics of collapsing system, which is defined as follows

\begin{equation}\label{md}
D_{R}=\frac{1}{R'}\frac{\partial}{\partial r}.
\end{equation}
Using Eqs.(\ref{m}) and (\ref{md}), we have
\begin{equation}\label{mr}
D_{R}m=\frac{l}{R'}\left[\frac{\dot{R}\dot{R'}}{4X^{2}}-\frac{\dot{R}^{2}X'}{4X^{3}}-\frac{R'R''}{4Y^{2}}+\frac{Y'R'^{2}}{4Y^{3}}+\frac{l\hat{Q}\hat{Q}'}{R}-\frac{l\hat{Q}^{2}R'}{2R^{2}}\right].
\end{equation}
Now Eqs.(\ref{d}), (\ref{q}) and (\ref{mr}), provide
\begin{equation}\label{mrr}
D_{R}m=2\pi Rl \left(4\mu+\frac{U}{\hat{E}}qY\right)+\frac{\hat{Q}\hat{Q}'l^2}{RR'}-\frac{\hat{Q}^{2}l^2}{R^{2}}.
\end{equation}
This expression yields the change in total energy contained inside the various cylindrical surfaces of different radii.
The term $4\mu +\frac{U}{\hat{E}}qY$, increases the energy as for the physically realistic fluid $\mu>0$ although it is affected
by the heat flux and $U<0$ reduces $\mu$. The second term implies the presence of electromagnetic field inside the gravitating source.
After the integration of Eq.(\ref{mrr}), we obtain
\begin{equation}\label{9}
m=\int^R_02\pi Rl\left(4\mu+\frac{U}{\hat{E}}qY\right)dR+\frac{\hat{Q}^2l^2}{2R}-\frac{l^2}{2}\int^R_0\frac{\hat{Q}^{2}}{R^{2}}dR.
\end{equation}
Here, we have assumed that $m(0)=0$.

Now we obtain $D_{t}U$, which is the acceleration of the collapsing matter inside the $\Sigma$.
From Eq.(\ref{92}), we get the following relation
\begin{equation}\label{9}
D_{t}U=\frac{1}{X}\frac{\partial}{\partial t}\left(\frac{\dot{R}}{X}\right)
\Rightarrow D_{t}U=\frac{\ddot{R}}{X^{2}}-\frac{\dot{R}\dot{X}}{X^{3}}.
\end{equation}
The above equation with Eq.(\ref{pr}), gives
\begin{equation}\label{93}
D_{t}U=-\left(\frac{m}{R^2}+8\pi (P_{r}-\xi\Theta )R\right)+ \frac{X'\hat{E}}{XY}+\frac{\hat{Q}^{2}}{R}\left(\frac{l^2}{2R^2}-1\right)+\frac{l}{8R^2}(1+U^2-\hat{E}^2).
\end{equation}
By the conservation law ($T^{\alpha\beta}_{;\beta}$=0), we deduce the following dynamical equations
\begin{eqnarray}\nonumber
P'_{r}&+&\frac{\dot{q}Y^{2}}{X}-(P_{\Theta}-P_{r})\frac{R'}{R}+\frac{qY^{2}}{X}\left(\frac{\dot{R}}{R}+\frac{3\dot{Y}}{Y}\right)+(P_{r}+\mu)\frac{X'}{X}
-\xi\Theta'\\\label{c1}&+&\frac{X'}{X}\xi\Theta+\left(\frac{\hat{Q}}{R^2}\right)\left(E'R-R'E\right)=0.
\end{eqnarray}
Using value of $\frac{X'}{X}$ from Eq.(\ref{c1}) into Eq.(\ref{93}) and considering field equations, after some algebra, we obtain
\begin{eqnarray}\nonumber
(P_{r}+\mu-\xi\Theta)D_{t}U&=&-\left(\mu+P_{r}-\xi\Theta\right)\left[\frac{m}{R^2}+8\pi R(P_{r}-\xi\Theta)+\frac{l^2\hat{Q}^{2}}{2R^{3}}-\frac{\hat{Q}^{2}}{R}+\frac{lU^2}{8R^2}\right]\\\nonumber
&-&\hat{E^{2}}\left[\frac{P_{r}}{R}-\frac{P_{\theta}}{R}-\frac{\hat{Q}^2}{\pi R^{3}}+\frac{l(\mu+P_{r}-\xi\Theta)}{8R^2}\right]\\\label{D1}&-&
\hat{E}\left[\frac{P'_{r}}{Y}+\frac{3q\dot{Y}}{X}+YD_{t}q+qY\frac{\dot{R}}{XR}-\frac{\xi\Theta'}{Y}+\frac{\hat{Q}\hat{Q}'}{2\pi YR^{2}}\right].
\end{eqnarray}
The factor $(P_{r}+\mu-\xi\Theta)$ being multiple of acceleration $D_T U$ plays the role of effective inertial mass density while same factor on the right hand side before the square bracket is the passive gravitational mass density. This factor is affected by the radial pressure and bulk viscosity but it is independent of electric charge. The first square bracket on right hand side shows the effects of dissipation and charge on the dynamical process.
The second square bracket gives the effects of local anisotropy, electric charge and gravitational mass density.
In the last square bracket $P'_{r}$ is pressure gradient and the terms involving $q$, $\xi$ and $\hat{Q}$ explain the collective effects of dissipation and electromagnetic field on the hydrodynamics of the collapsing source. The consequences of $D_{t}q$, will be dealt in the next section by driving the heat transport equation and then performing the possible coupling of the dynamical equation with the resulting heat transport equation.

\section{\textbf{Heat Transport Equation}}

 As already mentioned in the introduction, we shall use a transport equation that comes from the
M$\ddot{u}$ller-Israel-Stewart \cite{I1}-\cite{I3} second order phenomenological theory for dissipative fluids (by neglecting the thermodynamics viscous/heat coupling coefficients). Since we have introduced the bulk viscosity in fluid source so, we have to take accordingly the full casual approach as discussed in \cite{Herrera2}, but for the sake of simplicity, we neglect the thermodynamics viscous/heat coupling coefficients and only take into account the only transportation of heat flux governed by Cattaneo type equation \cite{I12} (leading to a hyperbolic equation for the propagation of thermal perturbation). Thus according to \cite{h1, h2}, the transport equation for the heat flux is
\begin{equation}\label{9}
\tau h^{\alpha\beta}V^{\gamma}q_{\beta;\gamma}+q^{\alpha}=-Kh^{\alpha\beta}(T_{,\beta}+T{a_{\beta}})-\frac{1}{2}KT^{2}{\left(\frac{\tau V^{\beta}}{KT^{2}}\right)}_{;\beta} q^{\alpha}.
\end{equation}
In the above equation $h^{\alpha\beta}$ denotes the projection onto the space orthogonal to $V^{\beta}$, $K$ is the thermal conductivity and $T$ and $\tau$
are temperature and relaxation time respectively. With the symmetry of the given interior spacetime heat transport equation has following form of independent component
\begin{eqnarray}\nonumber
YD_{t}q&=&-\frac{KYT'}{\tau}-\frac{KTY}{\tau}\left(\frac{X'}{X}\right)-\frac{1}{2\tau}KT^{2}Y^{3}{\left(\frac{\tau}{KT^2}\right)}^{\dot{}}q\\\label{23t}
&+&\frac{\dot{Y}q}{X}-\frac{Yq}{X\tau}-\frac{3}{2X}\dot{Y}qY.
\end{eqnarray}
Applying the value of $\frac{X'}{X}$ from Eq.(\ref{93}) in the above equation, we get
\begin{eqnarray}\nonumber
YD_{t}q&=&-\frac{KYT'}{\tau}-\frac{KTY^2}{\tau \hat{E}}D_{t}U-\frac{KTY^{2}}{\tau \hat{E}}\left(\frac{m}{R^2}+8\pi R(P_r-\xi\theta) -\frac{\hat{Q}^{2}l^2}{2R^3}\right.\\\nonumber &+&\left.\frac{\hat{Q}^{2}}{R}+\frac{l(1+U^2)}{8R^2}\right)-KTY^2\left(\frac{l \hat{E}}{8\tau R^2}\right)
-\frac{1}{2\tau}KT^{2}Y^{3}{\left(\frac{\tau}{KT^2}\right)}^{\dot{}}q\\\label{25}&+&\frac{\dot{Y}q}{X}-\frac{{Y}q}{\tau X}-\frac{3}{2X}\dot{Y}qY.
\end{eqnarray}
After substituting the value of $YD_{t}q$ in Eq.(\ref{D1}), we have
\begin{eqnarray}\nonumber
\left((P_{r}+\mu-\xi\Theta)-\frac{KTY^{2}}{\tau}\right)D_{t}U&=&-(P_{r}+\mu-\xi\Theta)\left(\frac{m}{R^2}+8\pi R(P_r-\xi\Theta) -\frac{\hat{Q}^{2}l^2}{2R^3}\right.\\\nonumber &+&\left.\frac{\hat{Q}^{2}}{R}+\frac{l(1+U^2)}{8R^2}\right)\left(1-\frac{KTY^{2}}{\tau(P_{r}+\mu-\xi\Theta)}\right)\\\nonumber
&-&\hat{E^{2}}\left[\frac{P_{r}-P_{\theta}}{R}-\frac{\hat{Q}^{2}}{\pi R^{3}}+\frac{l(P_{r}+\mu-\xi\Theta)}{8R^2}+\frac{lKTY^{2}}{8R^2\tau}\right]\\\nonumber
&-&\hat{E}\left[\frac{P'_{r}}{Y}+\frac{4q\dot{Y}}{X}-\frac{KYT'}{\tau}-\frac{1}{2\tau}KT^{2}Y^{3}{\left(\frac{\tau}{KT^2}\right)}^{\dot{}}q \right.\\ &-&\left.\frac{qY}{X\tau}-\frac{3q\dot{Y}Y}{2X}\right] \label{24t}.
\end{eqnarray}
Equation(\ref{24t}) may be written as
\begin{eqnarray}\nonumber
(P_{r}+\mu-\xi\Theta)\left(1-\alpha\right)D_{t}U&=&F_{grav}\left(1-\alpha\right)+F_{hyd}-\hat{E}\left[\frac{P'_{r}}{Y}+\frac{4q\dot{Y}}{X}-\frac{KYT'}{\tau}\right.\\ &-&\left.\frac{1}{2\tau}KT^{2}Y^{3}{\left(\frac{\tau}{KT^2}\right)}^{\dot{}}q -\frac{qY}{X\tau}-\frac{3q\dot{Y}Y}{2X}\right].\label{28t}
\end{eqnarray}
Here, $F_{grav}$, $F_{hyd}$ and $\alpha$ defined by

\begin{eqnarray}\nonumber
F_{grav}&=&-(P_{r}+\mu-\xi\Theta)\left(\frac{m}{R^2}+8\pi R(P_r-\xi\Theta) -\frac{\hat{Q}^{2}l^2}{2R^3}+\frac{\hat{Q}^{2}}{R}+\frac{l(1+U^2)}{8R^2}\right),\\
F_{hyd}&=&-\hat{E^{2}}\left[\frac{P_{r}}{R}-\frac{P_{\theta}}{R}-\frac{\hat{Q}^2}{\pi R^{3}}+\frac{l(\mu+P_{r}-\xi\Theta)}{8R^2}\right],\\
\alpha&=&\frac{KTY^{2}}{\tau(P_{r}+\mu-\xi\Theta)}.
\end{eqnarray}
From the above final resulting equation (\ref{28t}), it is noted how dissipation affects the final stage of charged collapsing cylinder. This fact was investigated for the first time in \cite{Herra1}, when the authors discussed the thermal conduction in systems out of hydrostatic equilibrium. They analyzed that the evolution of the gravitating source depend on parameter $\alpha$ (which is defined in term of thermodynamic variables), further for the validity of casuality, the constraints on $\alpha$ have been determined in that work.

It is clear that the left hand side of Eq. (\ref{28t}) will be zero as $\alpha\rightarrow1$, it confirm that the effective inertial mass density of fluid element tends to zero. Further, we observe that the inertial mass will be decreased as $\alpha$ exceeds than $1$. Moreover, $F_{grav}$ being a multiple of $(1-\alpha)$ is affected by this factor. Also, it is evident that both inertial mass and gravitational attraction are affected by the same factor $(1-\alpha)$. In other words, we can say that this equation satisfies the equivalence principle and we would like to point out that the factor $(1-\alpha)$ has no effects on $F_{hyd}$.
One may observe that a collapsing cylinder would evolves in such a way that the value of $\alpha$ keeps on increasing and attains a critical value of $1$. With the passage of time during collapse the rapid decrease in the force of gravity, may gradually results to alter the physical effects of right hand side
of equation (\ref{28t}). As it is clear from the definition of $\alpha$ that it is inversely related to effective inertial mass density, so as long as $\alpha$ increases from $1$ then there would be decrease in inertial mass density. Physically, it is only possible when gravitating source depicts the bouncing behavior. The factor $(1-\alpha)$ does not depend on charge parameter but it heavily depends on bulk viscosity which is explicitly clear from from Eq.(32). Further, one can see the dependence of factor $(1-\alpha)$ on the dissipative variables when a full casual approach \cite{Herrera2} is used to discuss the dynamics of dissipative collapse (see Eq.(54) of \cite{Herrera2}).

\section{Conclusion}
The cylindrically symmetric systems which combine translation along axis are exactly well-known to the general relativists. The study of such systems started by Weyl \cite{W1} and Levi-Civita \cite{L1} in the early of the 20th century immediately after the Einstein's theory of relativity. In the beginning physicists were  interested to find the gravitating objects that are exactly axially symmetric. The realistic fluids are very important in the modeling of astronomical objects. So, one can not ignore the effects of dissipation during the gravitational collapse.


Here, we discuss the gravitational collapse of charged radiating cylindrically symmetric stars. To this end, we formulated the Einstein field equations and conservation equation for non-static charged bulk viscous heat conducting anisotropic cylindrically symmetric source. Using the Misner and Sharp formalism, the dynamical equations are derived. Further, we have considered the two types of dissipation processes, heat dissipation associated to the radial heat flux and bulk viscosity. In order to see the effects of these terms on the dynamics of the collapse, we have excluded thermodynamics viscous/heat coupling coefficients in the heat transport equations in the context of M$\ddot{u}$ller-Israel-Stewart theory \cite{I1}-\cite{I3}.

The inclusion of the bulk viscosity in the fluid implies that we are assuming the relativistic Stokes equations, which corresponds to the irreversible thermodynamics. This equation does not satisfy the casuality, because implicitly it is assumed that corresponding bulk viscosity relaxation times vanishes. But this assumption is sensible, because within some approximations, such relation times could be neglected. A full casual approach to the dynamics of dissipative collapse has been analyzed with significant consequences in \cite{Herrera2} without excluding the thermodynamics viscous/heat coupling coefficients for the heat flux and bulk viscosity. As an implication of their analysis to astrophysical scenario, they pointed that in a pre-supernovae event, the dissipative parameters (particularly thermal conductivity) would be so large to produce a significant decreasing in the force of gravity which leads to the reversal of the collapse. We would like to mention that we have have introduced the bulk viscosity by the standard (non-casual) irreversible thermodynamics approach, so we have used the partially casual approach (the thermodynamics viscous/heat coupling coefficients have been excluded) to discuss the dynamics of considered dissipative source. Further, the form of Eqs.(17, 23, 24, 25,29,30,32) depends on the standard (non-casual) irreversible thermodynamics approach, which we have used in the present analysis. If one consider the full casual approach to discuss the dynamics of dissipative gradational collapse as in \cite{Herrera2}, then in the present case the term $-\xi\Theta$ in Eqs. (17, 23, 24, 25,29,30,32), will be replace by a dissipative variable $\Pi$.

Finally, from our analysis, it has been investigated that during the evolution of cylindrical star
charge, bulk viscosity and anisotropic stresses reduces the energy of the system and we conclude that

\begin{itemize}
\item The out flow of heat from the collapsing cylindrical star reduces the total energy of the system
\item Bulk viscosity reduces the radial pressure of the collapsing fluid
\item The bulk viscosity and charge of the fluid would affects the rate of collapse prominently
\item Active/passive gravitational mass density is affected by bulk viscosity and it is independent of electromagnetic field
\item In Eq.(\ref{28t}), the factor $(1 -\alpha)$ would explain the possible evolutionary stages of the charged dissipative cylinder
\item The term $\alpha$ is inversely related to gravitational mass density which is affected by bulk viscosity while it is linearly related to temperature to the fluid
\item The inclusion of bulk viscosity would increase the value of $\alpha$
\item For  $\alpha<1$, $\alpha>1$ and $\alpha=1$, we have expanding, collapsing and bouncing behavior of the fluid distribution, respectively.
\end{itemize}
\section{Acknowledgement}

The constructive comments and suggestions of anonymous referee are highly acknowledged.
 
\end{document}